\documentclass[]{raa}
\usepackage{graphicx,times}
\usepackage{natbib}
\usepackage{amssymb,amsmath}
\bibpunct{(}{)}{;}{a}{}{,}

\def\apj{ApJ}
\def\aap{A\&A}
\def\mnras{MNRAS}
\def\apjs{ApJS}
\def\apjl{ApJL}

\usepackage[a4paper=true,dvipdfm=true,pagebackref=true]{hyperref}
\hypersetup{pdftitle = The title of my PDF, pdfauthor = My name, pdfsubject= The subject, pdfkeywords = keyword1 keyword2 keyword3}
\hypersetup{colorlinks = true, linkcolor = green, anchorcolor = red, citecolor = blue, filecolor = red, pagecolor = red, urlcolor = red}

\begin{document}

   \title{Optical quasi-periodic oscillation and  color behavior of blazar PKS 2155-304
$^*$
\footnotetext{\small $*$ Supported by the National Natural Science Foundation of China.}
}

 \volnopage{ {\bf 2012} Vol.\ {\bf X} No. {\bf XX}, 000--000}
   \setcounter{page}{1}

   \author{Bing-Kai Zhang\inst{1}, Xiao-Yun Zhao\inst{1}, Chun-Xiao Wang\inst{1}, Ben-Zhong Dai
      \inst{2}
   }

   \institute{ Department of Physics, Fuyang Normal College, Fuyang
              236041, China; {\it zhangbk@ihep.ac.cn}\\
        \and
              Department of Physics, Yunnan University, Kunming 650091,
             China\\
	%\and
       %      Institue of High Energy Physics, Chinese Academy of  Sciences, Beijing 100049,
          %   China\\
\vs \no
   {\small Received 2014 February 2 ; accepted 2014 May 26}
}

\abstract{
PKS 2155-304 is a well studied BL Lac object in the southern sky. The historical optical data during different period have been collected and compiled. Light curves with a time span of 35 years have been constructed.  The $R$-band light curve has been analyzed by means of three methods: epoch folding method, Jurkevich method and discrete correlation function (DCF) method. It is derived that there is an evident periodic component of 317 days (i.e. 0.87 yr) superposed on a long-term trend with large-amplitude variation in the light curve. The variability of this source is accompanied by a slight color variation, and the brightness and color index are correlated with each other.  On the long time-scale, PKS 2155-304 exhibits a tendency of bluer-when-brighter, which means the spectrum becomes flatter when the source brightens.
}

   \authorrunning{B.-K. Zhang et al. }            %author_head in even pages
   \titlerunning{Optical quasi-periodic oscillation and  color behavior of PKS 2155-304}  % title_head in odd pages
   \maketitle

%________________________________________________ sections below
%
\section{Introduction}           %% first-level sections will be auto-capitalized
\label{sect:intro}

Blazars are the most extreme class of AGNs exhibiting very violent high energy phenomena.  The subset of blazars can be grouped into two very different categories: BL Lacertae-type objects (BL Lacs) and flat spectrum radio quasars (FSRQs).  They have been observed at all wavelengths from radio through very-high energy (VHE) gamma-rays, and exhibited rapid variability at all wavelengths on various time scales, from years and a few months to even shorter than an hour in some cases. The emission from blazars is thought to be highly beamed through the relativistic jet. Flux variation study is considered to be a powerful tool for understanding the structure and emission mechanism of AGNs. Periodical variabilities on a wide range of timescales have been reported by some investigations, e.g. \citep{fan00,xie08,king13}. The most notable case is that of OJ 287, which has an over 120-year-long light curve and a convincing 11-12 year periodicity \citep{kidger92,valtonen06}.  However, there is still some debate over whether blazar variability is periodic.

Multi-wavelength observations of blazars have been performed  for many years. They can give the general shape of the energy spectrum (i.e. flux versus frequency) which follows a power law proximately.  The energy spectrum can provide insight into the nature of the emission process. The low energy component between radio and optical even to X-ray frequencies,  is mainly attributed to synchrotron emission from non-thermal electrons in a relativistic jet. The high energy emission between X-rays and TeV $\gamma$-rays,  may be due either to the Compton up-scattering of low energy radiation by the synchrotron-emitting electrons \citep{botter07} or hadronic processes initiated by relativistic protons co-accelerated with the electrons \citep{mucke03}. Measurements of blazar spectral variability are important tools in constraining physical models. In the optical domain, the color index is often used to represent spectrum index, and its variations usually accompanies flux variabilities with different behaviours, such as "bluer when brighter" or "redder when brighter" trend.

The high-frequency peaked BL Lac PKS 2155-304 with a redshift of z = 0.116 is one of the brightest BL Lacs in the X-ray and EUV bands \citep{giommi98}. It was classified as a TeV blazar by the detection of VHE gamma rays by the Durham MK 6 telescopes \citep{chadwick99} , and then was confirmed by the H.E.S.S. collaboration with a high significant of 45$\sigma$ at energies greater than 160 GeV \citep{aharonian05}. This source has been observed on diverse timescales over a wide range of frequencies from radio to VHE gamma-rays, and shown rapid and strong variability \citep{dominici06,dolcini07, aharonian07,abramowski10, kastendieck11,abramowski12, aleksic12, barkov12}.
The long-term optical variability of PKS 2155-304 has been studied by some authors \citep{fan00, zhang96, kastendieck11}. \cite{fan00} investigated the periodic variations in the long-term optical light curves, found two possible periodicities of 3.7 years and 7.1 years. The short-term color varies complicatedly,  and shows different behaviours at different epoches, e.g. \citep{carini92,pesce97,xie99,dolcini07,bonning12}.

This research is aimed to investigate the periodic variations and long-term color behaviour with  the historical optical light curves of PKS 2155-304. The paper is organized as follows: In Section 2, the optical data are compiled and the historical light curve is constructed; In Section 3,  the technique used for searching periodicity are described and the periodic variations are investigated; In Section 4, the long-term color behaviour is studied; and then the discussion and conclusions are given in Section 5.

\section{Optical data}
All available historical archival data of PKS 2155-304 have been collected from the following literatures:
 \cite{griffiths79}, \cite{miller83}, \cite{brindle86}, \cite{hamuy87}, \cite{treves89}, \cite{mead90},
\cite{smith91}, \cite{carini92}, \cite{smith92}, \cite{jannuzi93}, \cite{urry93}, \cite{courvoisier95},
\cite{xie96}, \cite{heidt97}, \cite{pesce97}, \cite{bai98}, \cite{xie99}, \cite{bertone00},
\cite{tommasi01}, \cite{xie01}, \cite{gupta02}, \cite{dolcini07}, \cite{osterman07}, \cite{kastendieck11}. Up-to-date SMART optical data \citep{bonning12} have also been collected.  In sum, 179, 759, 1382, 8674 and 590 data points in $U$, $B$, $V$, $R$ and $I$ bands are compiled, respectively.
The observations cover the time duration from 1979 to 2013, and the time interval is about 35 yr.
The mean magnitudes in $U$, $B$, $V$, $R$ and $I$ bands are 12.52 $\pm$ 0.39, 13.54 $\pm$ 0.46, 13.04 $\pm$ 0.42, 12.78 $\pm$ 0.40, 12.23 $\pm$ 0.28, respectively.    The source shows violent activity. The amplitudes of the observed variability are $\Delta U$ = 1.50 mag, $\Delta B$ = 2.23 mag, $\Delta V$ = 2.57mag, $\Delta R$ = 2.40 mag, $\Delta I$ = 2.34 mag, respectively. The variations in different passbands show the similar trend. The light curves are well correlated with each other. The $R$ band light curves are displayed in two panels of Figure 1, so as to exhibit more details.

\begin{figure}
   \centering
  \includegraphics[width=10.5cm, angle=0]{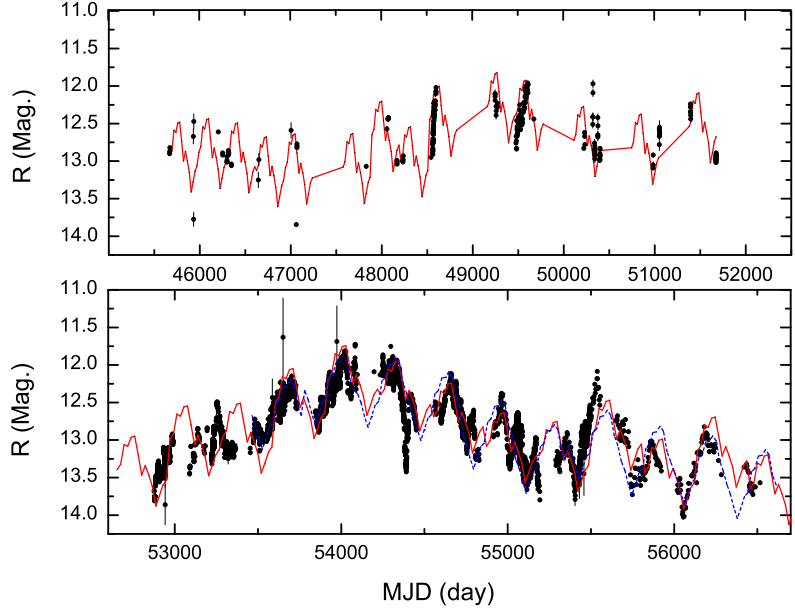}
  % \begin{minipage}[]{85mm}
   \caption{The historical optical $R$-band light curve of PKS 2155-304. The solid line and the short dash line are the folded light curves with a  317-day period using all the data and those after MJD 53400, respectively.}
%\end{minipage}
   \label{lightcurve}
   \end{figure}

It is clear that there are more intensive observations after MJD 52500. In Figure 1, one can see that the source exhibited oscillations with $\sim$0.7 mag throughout the duration of MJD 52500 - 54100, superposed on a general and slow brightening trend with a total magnitude of $\sim$2.  Then the source began to fade slowly till MJD 55200 by a total of $\sim$2 magnitude, accompanied by some small flickering with $\sim$0.7 mag. Subsequently, a large and sharp outburst appeared from MJD 55200 to 55300, after that the source brightness decreased with some small amplitude fluctuations.   It is obvious that the short term variations are superposed on the long-term and large variations.

\section{Periodicity analysis}

To search for periodic variations in the long-term light curve, epoch-folding technique has been employed.  This technique was described by \cite{leahy83} and extended by \cite{davies90,davies91}.  It is less affected by harmonics, and more effective for nonsinusoidal modulations than the Fourier transform technique which is most sensitive to a periodic component of sinusoidal form in shape. This method is also well suited for sparse and unevenly sampled data set. The $N$ data points of a time series are folded on a trial period and binned by phase. For the $i$th of $M$ phase bins, the mean $x_{i}$ and sample variance $\sigma_ {i}^{2}$ are computed, as is the overall mean, $<x>$. Then the $Q^{2}$ statistic is computed,
\begin{equation}
Q^{2}=\sum_{i=1}^{M}\frac{(x_{i}-<x>)^{2}}{\sigma_{i}^{2}},
\end{equation}
 which is distributed similarly to the $\chi^{2}$ statistic. \cite{davies90} pointed out that the test statistic is valid only for large sample sizes. He proposed an improved method for detecting periodicities based on the $L$-statistic, giving a greater sensitivity when the number of data points is limited. The $L$-statistic is  defined as,
\begin{equation}
L=\frac{(N-M)Q^{2}}{(M-1)(N-1)-Q^2},
\end{equation}
  which obeys an $F$ distribution with $M-1$ and $N-1$ degrees of freedom. The large value of  $L$ means that the corresponding trial period may be a true period in the light curve. It can be tested by calculating the false-alarm probability (the confidence with which one can reject the null hypothesis, i.e. no periodic component).

\begin{figure}
   \centering
  \includegraphics[width=10.5cm, angle=0]{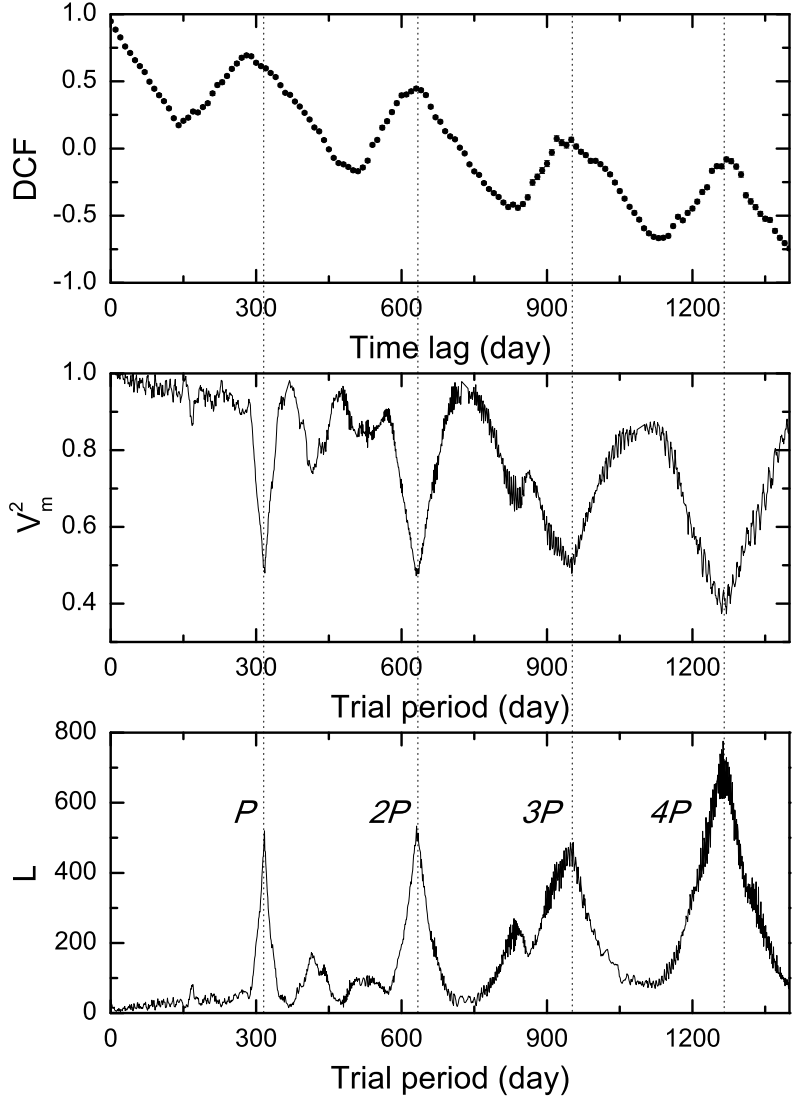}
  % \begin{minipage}[]{85mm}
   \caption{Upper: $DCF$ as a function of time delay, $\tau$; middle: $V^{2}_{m}$ as a function of trial periods; lower: $L$-statistic as a function of trial periods for the optical R-band light curve. The vertical dotted lines are drawn to guide the eyes.}
%\end{minipage}
   \label{period}
   \end{figure}

The $R$-band observations are used to search for periodicity because there are more data available in the $R$-band than others.  For each trial period, the data are folded into 20 phase bins, and then the $L$-statistic are calculated for a range of trial periods from 1 d to 1400 d in steps of 1 d.  The $L$ is plotted as a function of the trial period in the lower panel of Figure~\ref{period}. The strong peak at 317 $\pm$ 12 d (i.e. 0.87 yr), is clearly visible (the error corresponds to the half of the full width at half maximum (FWHM)), which means the periodicity $P$=317 d. The integer multiples of this period are also detected strongly at 631 d, 954 d and 1263 d which correspond to 2$P$, 3$P$ and 4$P$, respectively.

 To confirm the periodicity of 317 d,  Jurkevich method \citep{jurkevich71} and discrete correlation function (DCF) method \citep{edelson88} are also applied to search for periodic signals in the light curve. The Jurkevich method adopted the folding technique, while it is based on the expected mean square deviation. The light curve is folded according to test periods and then all data are grouped to $20$ phase bins. The sum variance $V^{2}_{m}$ = $\sum_{i=1}^{20} $$V_{i}^{2}$ of all phase are computed and shown in the middle panel of Figure~\ref{period}. The minimum of $V^{2}_{m}$ suggests that the trial period may be true one. Obvious dips locate at 318 d, 630 d, 951 d and 1270 d, which represents the principal period, $P$, and its integer multiples, respectively. They are consistent well with those detected by epoch-folding method.

 The DCF method is an useful tools to investigate not only correlations between different light curves but also periodic component in a light curve. For two time series, $a$ and $b$,  $DCF(\tau)$ is defined as  the mean of $(a_{i}-\overline{a}) (b_{i}-\overline{b}) /\sigma_{a}\sigma_{b}$ for all pairs with $\tau - \Delta \tau/2\leq t_{j}-t_{i}<\tau + \Delta\tau/2$. When $a=b$, the position of $DCF$ peak gives the periodicity information in the light curve. The $DCF$ for the $R$ band is computed and plotted as a function of $\tau$ (the upper panel of Figure~\ref{period}).  The peak at $P=280$ d can be seen obviously. It is a little different to the period of 317 d derived by the epoch-folding technique.  The subsequent peaks appear at 630 d, 920 d and 1270 d, which are near to the corresponding values detected by the other two methods. The last two peaks are very low, which means that they are modulated by a longer-term variation trend.

\section{Color behavior}
\begin{figure}
   \centering
  \includegraphics[width=10.5cm, angle=0]{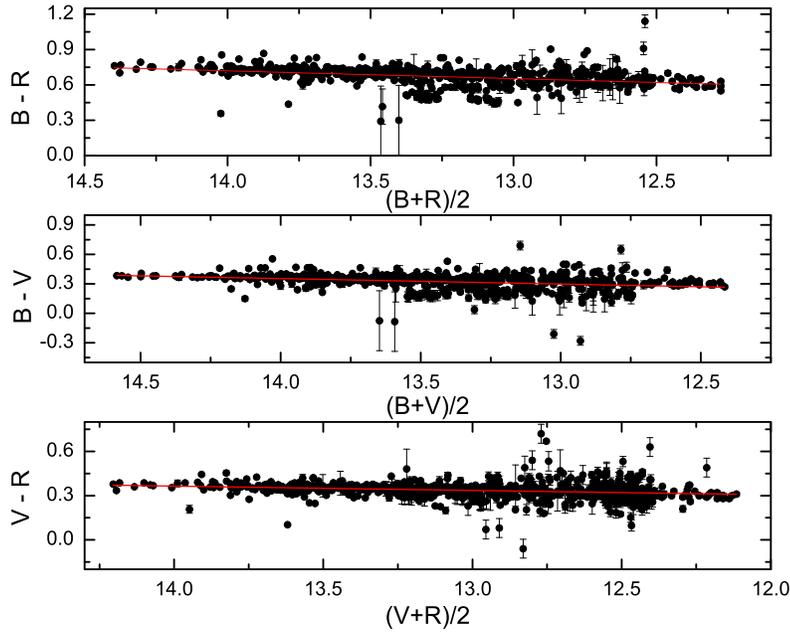}
  % \begin{minipage}[]{85mm}
   \caption{Color indices versus magnitude of PKS 2155-304. The solid lines represent the best linear fitting.}
%\end{minipage}
   \label{color}
   \end{figure}

The behavior of colors provides a clue to understand the mechanism of time variations in blazars.
 Since the blazars varies rapidly, the color index should be ideally evaluated using simultaneous observations in the different bands.  However, there are few exactly simultaneous data. The color indices with quasi-simultaneous data have been calculated. The mean values of color indices are listed in Column 2 of Table~\ref{regression}. The correlation between the color and magnitude is investigated. Figure \ref{color} displays the color indices versus magnitudes. Linear regression analysis is performed, and the results are listed in Table~\ref{regression}. Columns 3 to 6 list the slope, the correlation coefficient $r$, chance probability $Prob.$ and number $N$, respectively.

\begin{table}
\bc
%\begin{minipage}[]{100mm}
\caption[]{The results of linear regression analysis between color index and magnitude.
\label{regression}}
%\end{minipage}
\setlength{\tabcolsep}{10pt}
\small
 \begin{tabular}{cccccc}
  \hline\noalign{\smallskip}
Color & mean & slope & $r$ & $Prob.$ & $N$ \\
  \hline\noalign{\smallskip}
$B-R$ & 0.67 $\pm$ 0.09 & 0.066 $\pm$ 0.008 & 0.33 & $<10^{15}$ & 635 \\
$B-V$ &  0.32 $\pm$ 0.08 &  0.056 $\pm$ 0.007 & 0.30 & 3.56$\times 10^{-15}$ & 678 \\
$V-R$ &  0.33 $\pm$ 0.06 &  0.029 $\pm$ 0.004 & 0.23 & 2.51$\times 10^{-11}$ & 837 \\
  \noalign{\smallskip}\hline
\end{tabular}
\ec
%% place \tablecomments and \tablerefs below \end{center| and \end{center}:
%% you may leave the table-width parameter to editors or set to your actual size
\end{table}

According to the results listed in Table~\ref{regression}, one can see that data points in Figure~\ref{color} are fitted well by straight lines. The small correlation coefficients suggest that there is a weak correlation between color index and magnitude.
There is a general indication that the object becomes bluer when it is brighter.

\section{Discussion and conclusions}
Three methods are applied to search for periodic component in the $R$-band light curve. The results are consistent well with each other. A 317-day (i.e. 0.87 yr) periodic oscillation can be seen clearly. For the periodicity of 317 d, the corresponding $L$ is  519. This means that the chance probability of obtaining such a large value is extremely small. According to the $F$-distribution, the chance probability with the null hypothesis is less than $10^{-20}$. So, 317 day is suggested to be a true periodicity. In addition, this value can be confirmed by Jurkevich method.  \cite{kidger92} provided a fraction $f=(1-V^{2}_{m})/V^{2}_{m}$ to assess the significance of the periodicity derived by Jurkevich method. A value of $f\geq0.5$ implies a strong periodicity in the light curve. In this analysis, $f=1.2$ suggests 317-day periodicity is true.

\begin{figure}
   \centering
  \includegraphics[width=10.5cm, angle=0]{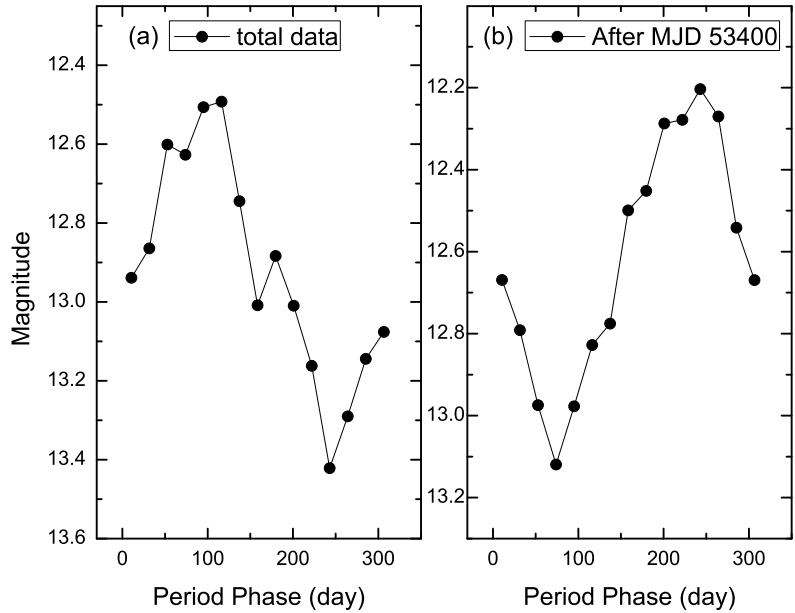}
  % \begin{minipage}[]{85mm}
   \caption{The folded light curves with the period of 317 days. (a): All the data are employed; (b): The data after MJD 53400 are employed. The curves seem to be in compliance with a sinusoidal wave.}
%\end{minipage}
   \label{phase-period}
   \end{figure}

 In Figure~\ref{lightcurve}, from MJD 53500 forwards, one can see five obvious peaks well separated by nearly 317-day intervals.  The data shows the average separation is 316 days among these five peaks which are spaced by 321 day, 314 day, 325 day and 304 day, respectively. Going backwards and forwards from these five peaks, the periodicity seems to disappear.  But after MJD 55500, the periodicity appears again with 328-day and 315-day intervals. Sometimes, the positions of the peaks are difficult to identified exactly. So, the folded light curve is calculated with the period of 317 days (Figure~\ref{phase-period}), and  superimposed on the light curve in Figure~\ref{lightcurve} (the general shape of the whole light curve is taken into consideration).   One can see that, except the peaks near MJD 53250 and MJD 55060, the folded light curves are consistent well with most of the peaks and the dips in the light curve, but nevertheless, the last three peaks seem to not be in the same periodic sequence with the previous ones.  So 317 day is not a strict and persistent periodicity in the light curve. The variations with time scale longer than 317 days can also be seen from Figure~\ref{lightcurve}. This tells us that in the light curve there may be a longer periodic component which is difficult to be dug out at present.

For PKS 2155-304, a 5-day roughly continuous light curve in November 1991  appeared to have a quasi-periodicity of 0.7 day \citep{urry93}.  However, this periodicity was not confirmed by a more rigorous analysis, although more data covering the whole November 1991 were adopted \citep{edelson95}. Using the historical data from 1977 to 1994, \cite{fan00} found two possible periodicities of 4.16 years and 7.0 years in the $V$ band light curve. The two possible periodicities are respectively 5 and 8 times the periodicity (0.87 year) found in this work.  They also pointed out the hints of period less than 4.0 years which could not confirmed at that time due to the limitation of data sample. \cite{kastendieck11} characterised the optical behaviour of PKS 2155-304 with a long-term light curve, however, they found no clear evidence for periodic behavior on any timescales.

Several models have been developed to explain the periodic variations, for example supermassive binary black hole model\citep{valtaoja00,xie02} and helical jet model (see \cite{rani09} and references therein). The quasi-periodic variations seem to suggest that there is a supermassive binary black hole (SMBBH) system in the center of a source.
The periodic brightness fading is caused by the eclipse of the system.  In the case of PKS 2155-304, the emission is dominated by synchrotron emission and inverse Compton emission which both arise from relativistic jets.   The quasi periodic variations are mostly caused by blobs propagating in a helical jet.  The effect is similar to that of a jet whose direction changes.  When the viewing angle is small, the source brighter, and vice versa. In this case, one would expect to see a rather exact variability pattern for some periods until the blob vanishes, and then a repetition with a new blob. From Figure~\ref{lightcurve}, the light curve seems to have a possible 5-peak stretch of almost periodic variations with a periodicity of 317 days, a break in the pattern, and then a 3-peak repetition of the 317-day periodicity, again followed by a break. These phenomena are consistent well with the expectation.

%\subsection{color}
In general, there are five kinds of color behaviors:
(1) bluer-when-brighter (BWB) in the whole data sets. It is a well-observed feature in blazars especially in BL Lac object  \citep{gu06,rani10,zhang10,ikejiri11}. This trend is generated by a variation component with a constant and relatively blue color and an underlying red component \citep{ikejiri11};
(2) redder-when-brighter (RWB) in the whole data sets. Most of FSRQs follow this redder-when-brighter tendency, which suggests the presence of a steady blue accretion disk component underlying the more variable jet emission \citep{rani10,bonning12};
(3) cycles or loop-like pattern, e.g. S5 2007+777 and 3C 371 \citep{xilouris06}, S5 0716+714 \citep{wu07}, and OJ 287 \citep{bonning12}, which may be caused by different amplitude and time delay in different spectral bands;
(4) redder-when-brighter at the low state while bluer-when-brighter at the high state (RWB to BWB), e.g. AO 0235+164 \citep{bonning12}, PKS 0537-441 \citep{zhang13};
(5) stable when brighter (SWB) or no correlation with brightness in the whole  data sets \citep{ikejiri11,gu11}.

Since 1970s, color of PKS 2155-304 has been investigated by several authors. The color behavior showed very different and complex tendencies on different time scales and during different periods. On short time scales of days to months, during 1990 October and December, PKS 2155-304 showed a tendency to be bluer when the object was brighter (BWB)\citep{smith91}, but the tendency was not observed in 1990 November, the optical spectral index remained relatively constant even though the object brightened by nearly one magnitude (SWB) \citep{smith92}. Observations during 1991 November showed a constant spectral slope in the $U-I$ domain (SWB) \citep{courvoisier95}. During the period 1994, the average colors remained relatively constant, with no correlation with brightness (SWB) and slightly BWB \citep{pesce97}. During the campaign in 1995, there was clear evidence of hardening when the source got brighter (BWB) \citep{paltani97}. During 1996 August-October, bluer-when-brighter trend (BWB ) was observed by \cite{xie99}. From May to December 2005, no apparent correlation between spectral index and brightness (SWB) was found by \cite{dolcini07}, but the source exhibited a rather soft spectral shape during its high state (RWB).

  On long time scales of years, this source during 1979-1982 became redder as it brightened (RWB) \citep{miller83}. From 1983 to 1985, it was found that the
  higher state was harder than the lower one (BWB) \citep{treves89}. \cite{zhang96} collected the pre-1994 observations, and found no correlation between brightness and colors (SWB). Between 2001 and 2003, the optical colors showed a BWB phenomena but opposite ones in the infrared domain \citep{dominici06}. \cite{ikejiri11} found this source exhibited BWB trend during 2008 and 2010. The observations from 2008 to 2010, the overall trend over several years revealed no strong correlation between color and brightness (SWB) \citep{bonning12}. The data from April 2005 to June 2012 observed by REM telescope showed the color did not varied with the brightness (SWB) \citep{sandrinelli14}.

It is clear that PKS 0537-441 exhibited BWB or SWB trends most of time.  In our color-magnitude analysis, the data covering 35 years is used. On the longest time scale till now, the source shows a clear bluer-when-brighter (BWB) trend which means the spectrum hardens when the source brightens. \cite{rani10} suggested that BWB and RWB both can be accommodated within shock-in-jet models.

%\section{Conclustions}
In conclusion, the long-term possible periodic variations have been investigated, a 317-day (0.87 yr) periodic component in the light curve has been detected, and it is more convincing. On the whole, the source varies with a 0.87 yr periodicity superposed on a long-term slower trend. PKS 2155-304 shows complex color behaviour on different time scales. This analysis suggest a clear  bluer-when-brighter chromatism on the long-term time scale.

\normalem
\begin{acknowledgements}
We thank the referee for great helps. This work is supported by the National Natural Science Foundation of China  (11273008).

\end{acknowledgements}

\bibliographystyle{raa}
%\bibliography{bibtex}

\end{document}